\documentclass{JHEP3} % 10pt is ignored!

\JHEPspecialurl{http://jhep.sissa.it/JOURNAL/JHEP3.tar.gz}

\usepackage{epsfig,multicol,bbm,cite,amsmath}
\newcommand\fverb{\setbox\pippobox=\hbox\bgroup\verb}
\newcommand\fverbdo{\egroup\medskip\noindent%
            \fbox{\unhbox\pippobox}\ }
\newcommand\fverbit{\egroup\item[\fbox{\unhbox\pippobox}]}
\newbox\pippobox
\def\ifm#1{\relax\ifmmode#1\else$#1$\fi}

\makeatletter
\newdimen\z@ \z@=0pt % can be used both for 0pt and 0
\newskip\z@skip \z@skip=0pt plus0pt minus0pt
\def\m@th{\mathsurround=\z@}
\def\ialign{\everycr{}\tabskip\z@skip\halign} % initialized \halign
\def\eqalign#1{\null\,\vcenter{\openup\jot\m@th
  \ialign{\strut\hfil$\displaystyle{##}$&$\displaystyle{{}##}$\hfil
    \crcr#1\crcr}}\,}
\makeatother
\catcode`@=11 %%At signs are letters
\newdimen\z@ \z@=0pt % can be used both for 0pt and 0
\newskip\z@skip \z@skip=0pt plus0pt minus0pt
\def\m@th{\mathsurround=\z@}
\def\ialign{\everycr{}\tabskip\z@skip\halign} % initialized \halign
\def\eqalign#1{\null\,\vcenter{\openup\jot\m@th
  \ialign{\strut\hfil$\displaystyle{##}$&$\displaystyle{{}##}$\hfil
      \crcr#1\crcr}}\,}
\catcode`@=12 % at signs are no longer letters
\def\figb#1;#2;{\parbox{#2cm}{\epsfig{file=#1.eps,width=#2cm}}}
\let\cl=\centerline
\def\figbc#1;#2;{\cl{\figb #1;#2;}}
\newcommand{\eV}{{e\kern-.07em V}}
\title{\mathversion{bold}Measurement of the charged kaon lifetime with the KLOE detector}
\author{The KLOE collaboration:\\
F.~Ambrosino,$^{d,*}$
A.~Antonelli,$^a$
M.~Antonelli,$^a$
F.~Archilli,$^a$
C.~Bacci,$^g$
P.~Beltrame,$^d$
G.~Bencivenni,$^a$
S.~Bertolucci,$^a$
C.~Bini,$^e$
C.~Bloise,$^a$
S.~Bocchetta,$^g$
F.~Bossi,$^a$
P.~Branchini,$^g$
R.~Caloi,$^f$
P.~Campana,$^a$
G.~Capon,$^a$
T.~Capussela,$^a$
F.~Ceradini,$^g$
S.~Chi,$^a$
G.~Chiefari,$^d$
P.~Ciambrone,$^a$
E.~De~Lucia,$^a$
A.~De~Santis,$^f$
P.~De~Simone,$^a$
G.~De~Zorzi,$^f$
A.~Denig,$^b$
A.~Di~Domenico,$^f$
C.~Di~Donato,$^d$
B.~Di~Micco,$^g$
A.~Doria,$^d$
M.~Dreucci,$^a$
G.~Felici,$^a$
A.~Ferrari,$^a$
M.~L.~Ferrer,$^a$
S.~Fiore,$^f$
C.~Forti,$^a$
P.~Franzini,$^f$
C.~Gatti,$^a$
P.~Gauzzi,$^f$
S.~Giovannella,$^a$
E.~Gorini,$^c$
E.~Graziani,$^g$
W.~Kluge,$^b$
V.~Kulikov,$^j$
F.~Lacava,$^f$
G.~Lanfranchi,$^a$
J.~Lee-Franzini,$^{a,h}$
D.~Leone,$^b$
M.~Martini,$^a$
P.~Massarotti,$^{d,}$\footnote{
\rm{Corresponding authors.}\\ \it{e-mail addresses}\rm{:
  fabio.ambrosino@na.infn.it (F. Ambrosino), paolo.massarotti@na.infn.it
  (P. Massarotti)}}~~W.~Mei,$^a$
S.~Meola,$^d$
S.~Miscetti,$^a$
M.~Moulson,$^a$
S.~M\"uller,$^a$
F.~Murtas,$^a$
M.~Napolitano,$^d$
F.~Nguyen,$^g$
M.~Palutan,$^a$
E.~Pasqualucci,$^f$
A.~Passeri,$^g$
V.~Patera,$^{a,e}$
F.~Perfetto,$^d$
M.~Primavera,$^c$
P.~Santangelo,$^a$
G.~Saracino,$^d$
B.~Sciascia,$^a$
A.~Sciubba,$^{a,e}$
A.~Sibidanov,$^{a}$
T.~Spadaro,$^a$
M.~Testa,$^f$
L.~Tortora,$^g$
P.~Valente,$^f$
G.~Venanzoni,$^a$
R.Versaci,$^a$
G.~Xu,$^{a,i}$
\\
\llap{$^a$}Laboratori Nazionali di Frascati dell'INFN, Frascati, Italy\\
\llap{$^b$}Institut f\"ur Experimentelle Kernphysik, Universit\"at Karlsruhe, Germany\\
\llap{$^c$}Dipartimento di Fisica dell'Universit\`a e Sezione INFN, Lecce, Italy\\%%%                               h=>f
\llap{$^d$}Dipartimento di Scienze Fisiche dell'Universit\`a  ``Federico II'' e Sezione INFN, Napoli, Italy\\
\llap{$^e$}Dipartimento di Energetica dell'Universit\`a ``La Sapienza'', Roma, Italy\\%%%                       m=j
\llap{$^f$}Dipartimento di Fisica dell'Universit\`a ``La Sapienza'' e Sezione INFN, Roma, Italy\\
\llap{$^g$}Dipartimento di Fisica dell'Universit\`a ``Roma Tre'' e Sezione INFN, Roma, Italy\\
\llap{$^h$}Physics Department, State University of New York at Stony Brook, USA\\%%%                        l=>i
\llap{$^i$}Institute of High Energy Physics of Academia Sinica,  Beijing, China\\%%%                                j=>h
\llap{$^j$}Institute for Theoretical and Experimental Physics, Moscow, Russia\\%%%                                  i=>g
}
\preprint{pre-print xxx}
\received{\today}       %%
\revised{\today}
\accepted{\today}       %% These are for published papers.

\abstract{We have measured the charged kaon lifetime using a sample of 15
  $\times$ 10$^6$ tagged kaon decays. Charged kaons were produced in pairs
  at the DA$\Phi$NE $\phi$-factory, $e^+e^- \to \phi \to K^+ K^-$. The
  decay of a $K^+$ was tagged by the production of a $K^-$ and
  viceversa. The lifetime was obtained, for both charges, from independent
  measurements of the decay time and decay length distributions. From fits
  to the four distributions we find $\tau$ = (12.347$\pm$0.030) ns.}


\begin{document}
\section{Introduction}
While the lifetime of the $K_S$ is well measured and measurements of the
lifetime of the $K_L$ \cite{Gaia},\cite{Mario} have been recently
performed, the most precise measurement of the lifetime of the charged kaon
dates back to 1971: $\tau_{\pm}=(12.380\pm0.016)$ ns \cite{ott}. At the
time, the agreement with previous measurements was poor and more recent
measurements did not improve the agreement. The 2006 PDG \cite{PDG} gives
$\tau_{\pm}=(12.385\pm0.025)$ ns corresponding to an accuracy of
0.2$\%$. However the set of measurements upon which the above value is
based is not self consistent. The probability that the spread of the values
used by PDG be due to the statistical fluctuations is very low,
0.17$\%$. Averaging an inconsistent set is not a valid procedure. The PDG
in fact enlarges the errors by a factor of 2.1. It is important therefore to
confirm the value of $\tau_{\pm}$. We report in the following on new
measurements of $\tau_{\pm}$. The statistical error on the lifetime
obtained fitting a decay curve over a time interval $\Delta$t is given
by\footnote{this formula is just the standard maximum likelihood 
  estimator error evaluation for an exponential probability distribution
  function integrated over a finite time interval $\Delta$t}:
\begin{equation}
  \frac{\delta \tau}{\tau} = \frac{1}{\sqrt{N}}\times \frac{e^{T} -
  1}{\sqrt{1 + e^{2T} - e^{T}(2+T^2)}}
  \label{eq:error}
\end{equation}
where T = $\Delta$t/$\tau$ is the time interval in $K^{\pm}$-lifetime
units. With $T \approx 2$ and number of events $N \approx 4 \times 10^6$,
the best statistical accuracy reachable in our case is $0.1\%$, if no other
sources of statistical error were present. We have developed two different
methods, one employing the reconstruction of the kaon path length and the
other measuring directly the kaon decay time.
%%%%%%%%%%%%%%%%%%%%%%%%%%%%%%%%%%%%%%%%%%%%%%%%%%%%%%%%%%%%%%%%%%%%%%%%%%%%%%
\section{The KLOE detector}
Data were collected with the KLOE detector at DA$\Phi$NE \cite{DAFNE},
the Frascati $e^{+} e^{-}$ collider operated at a
center of mass energy $ W =M_{\phi}\sim$1020
MeV. Equal-energy positron and electron beams collide with a
crossing angle of $\pi - 25$ mrad, which results in a small transverse
momentum ( $p_{\phi}\sim$ 12.5 MeV) of the $\phi$ mesons produced. The beam
spot has $\sigma_x \approx$ 2 mm $\sigma_y \approx$ 0.02 mm  and $\sigma_z
\approx$ 30 mm. The
KLOE detector is inserted in a 0.52 T magnetic field. It consists of a
large cylindrical drift chamber (DC), surrounded by a fine sampling
lead-scintillating fibers calorimeter (EMC). The DC \cite{DC}, 4 m diameter
and 3.3 m long, has full stereo geometry and operates with a gas mixture of
90\% Helium and 10\% Isobutane. Momentum resolution for tracks with large
PT is $\sigma(p_{\bot})/p_{T}\sim 0.4\%$. Spatial resolution is
$\sigma_{r\phi} \sim150 \mu$m and $\sigma_{z}\sim$ 2 mm. Vertices are
reconstructed with a 3D accuracy of $\sim$3 mm. The DC inner
walls are made in carbonium fibers with thickness of $\sim$1 mm. The beam
pipe walls are made of AlBeMet, an alloy of beryl-aluminum 60$\%$-40$\%$, with
thickness of $\sim$0.5 mm. The EMC \cite{EMC}, divided into a barrel and
two endcaps, for a total of 88 modules, covers 98\% of the solid
angle. Arrival times of particles, 3D positions and the energy deposits are
obtained from readout of signals on both ends of the module, with a
granularity of $\sim$ 4.4 $\times$ 4.4 cm$^{2}$, for a total of 2240 cells
arranged in five layers. Cells close in time and space are grouped into a
calorimeter cluster. Resolution on energy and time measurement are $
\sigma_{E}/E = 5.7\%/\sqrt{E(\rm{GeV})}$ and $\sigma_{T} =
57\textrm{ps}/\sqrt{E(\rm{GeV})} \oplus 100\; \textrm{ps}$. The trigger
\cite{TRIGGER}, used for this analysis, is based on the coincidence of at
least two local energy deposits in the EMC, above a threshold of 50 MeV in
the barrel and 150 MeV in the end caps. Cosmic rays muons crossing the
detector are vetoed. Since the trigger time formation is larger than the
interbunch time ($\sim$ 2.7 ns), the trigger operates in continuous mode
and its signal is synchronized with the DA$\Phi$NE accelerator Radio
Frequency. The trigger $T_0$ for an event is thus displaced in time with
respect to the correct crossing by an integer multiple of the interbunch
time which is different event by event. The correct bunch crossing is found
after the event reconstruction. For this analysis a sample 3$\times 10^8$of
$\phi \to K^+ K^-$, events generated with the KLOE MonteCarlo
(MC)\cite{EVCL}, has been used.
%%%%%%%%%%%%%%%%%%%%%%%%%%%%%%%%%%%%%%%%%%%%%%%%%%%%%%%%%%%%%%%
\section{The tag procedure}
$\phi$-mesons decay, in their rest frame, into anti-collinear $K\bar{K}$ pairs.
In the laboratory this remains approximately true because of the small
$\phi$-meson $\beta$. The detection of a $K^{+}(K^{-})$ tags the presence
of a $K^{-}(K^{+})$ of given momentum and direction.
The decay products of the $K^{+} K^{-}$ pair define two spatially separated
regions called the tag and the signal hemispheres.
Identified $K^{\mp}$ decays tag a $K^{\pm}$ sample of known total count
$\mathcal{N}$. This procedure is a unique feature of a $\phi$-factory and
allows to measure absolute branching ratios. For this analysis charged
kaons are tagged using the $K^{\pm}\rightarrow
\mu^\pm\rlap{\raise1.2ex\hbox{\scriptsize($-$)}} \kern.3em\nu_{\,\mu}$
($K_{\mu2}$) decay. This decay corresponds to about 63\% of the charged
kaon decay width \cite{PDG} and since $BR(\phi \rightarrow K^+K^-)\simeq
49\%$ \cite{PDG} and $\sigma(e^+ e^- \to \phi) \sim $ 3 $\mu$b, there are
about $1.1 \times 10^6 K^+K^-$ events/pb$^{-1}$. The $K_{\mu2}$ decay
is clearly identified since it exhibits a peak,
reconstucted with a
resolution of about 1 MeV, at $\approx$ 236 MeV in the
momentum spectrum of the secondary tracks in the kaon rest frame. In order to
minimize biases to trigger efficiency, we ensure that the tagging kaon did
provide the trigger of the event by requiring a cluster in the EMC (associated to the muon track) which satisfies the trigger conditions. Hereafter these events are called
self-triggered tags. We find $N_{\rm selftrg\  tag} \approx 1.2 \times
10^5$ per pb$^{-1}$. The MC simulation shows that the contamination of the
selected $\phi \rightarrow K^+ K^-$ sample is negligible. The $K_{\mu 2}$
tag allows a precise determination of the correct bunch crossing of the
event using the muon and kaon
track lengths and the momenta measured in the DC and the arrival time of
the muon in the EMC. The tagging efficiency is almost independent on the
proper decay time of the signal kaon. The small residual correlation has
been evaluated with the Monte Carlo and checked with data using doubly
tagged events, the events in which both the $K^+$ and the $K^-$ are
reconstructed and tag the event.
%%%%%%%%%%%%%%%%%%%%%%%%%%%%%%%%%%%%%%%%%%%%%%%%%%%%%%%%%%%%%%%%%%%%%%%%
\section{Signal selection}
\noindent
The measurement is performed using data from an integrated luminosity
$\mathcal{L}$= 210 pb$^{-1}$ collected at the $\phi$ peak. The average
$\phi-$meson momentum and the coordinates of the $e^+e^-$ interaction point
(IP) are measured run by run with Bhabha scattering events. $K_{\mu 2}$
tags of both charges have been used. We developed two analysis methods: the
kaon decay length and the kaon decay time. The two methods have comparable
precision and different systematics; this allows a useful cross-check of
the results. We use a coordinate system where the x-axis points to the
center of the collider, the z-axis bisects the two beam lines and the
y-axis is vertical. For both methods the kaon decay vertex position
$(x,y,z)$ is searched for in a fiducial volume (FV) defined by:
\begin{equation}
  40\ \mbox{cm} < R < 150\ \mbox{cm}, \kern.5cm |z| \le 150\ \mbox{cm}
\label{eq:FV}
\end{equation}
where $R = \sqrt{x^2 + y^2}$.
\subsection{Kaon decay length method}
The measurement of the charged kaon decay length requires the
reconstruction of the kaon decay vertex using only DC-information. We use
any kaon decay mode. The evaluation of the vertex reconstruction efficiency
uses a data control sample obtained by EMC-information only. Given a
charged kaon fulfilling the self-triggering tag requirements, our signal is
given by the opposite charged kaon decaying in the FV. The signal kaon
track reconstructed in the DC must satisfy the following requests:
\begin{eqnarray}
  \nonumber
  && R_{\rm{PCA}} < 10\ \mbox{cm},\\
  && |z_{\rm{PCA}}| < 20\ \mbox{cm},\\
  \nonumber
  && 70 <p_K< 130\ \mbox{MeV}.
\end{eqnarray}
where $R_{\rm {PCA}} = \sqrt{x^2_{\rm{PCA}} + y^2_{\rm {PCA}}}$ and PCA
indicates the point of closest approach of the kaon track to the
interaction point. The low level of background allows to keep these cuts very
loose, even accounting for the effect of multiple scattering and energy
loss of the kaon in the DC wall and into the beam pipe. We ask the
reconstruction of a vertex formed by the kaon track and one of its charged
decays, lying inside the FV. The only source of background is given by
events in which one kaon track is split in two pieces by the reconstruction
algorithm so as to mimic a decay vertex. This happens for kaon with low
$p_t$ which describe cirles in the DC. In order to reject these events we
used the momentum of the charged secondary particle ($p^{\ast}$) evaluated
in the kaon rest frame, using the kaon mass hypothesis. With the cut
$p^{\ast}>100$ MeV we lose about 5.4\% of signal and we reduce this
background to the 0.46\% level. Since charged kaons have an average
velocity $\beta \approx$ 0.2 they lose a non negligible fraction of their
energy traversing the beam pipe and DC walls and in the DC gas. Therefore,
given the decay length, we have to correct for the corresponding change in
$\beta$, of the order of 25$\%$, to evaluate the proper time. Once the
decay vertex has been identified, the kaon track is extrapolated backwards
to the interaction point in 5 mm steps, $\Delta l_i$, taking into account
the average ionization energy loss $dE/dx$ to evaluate its velocity
$\beta_i$ in each step. The kaon proper decay time, $t^*$ is finally
evaluated as
\begin{equation}
t^* = \sum_i \Delta t_i =
\sum_i \frac{\sqrt{1-\beta^2_i}}{\beta_i} \frac{\Delta l_i}{c}
\label{eq:beta_ch}
\end{equation}
In the range $12 <t^*< 40$ ns we collected
$1.7\times10^{6}$ events. The average resolution in $t^*$, evaluated
from MC simulation, is of the order of 1 ns, see fig. \ref{fig:reso_paper}
and its effects are taken into account in the fit procedure by means of
a resolution smearing matrix, as described in section \ref{sec:fit}, which
makes the resolution folding.
\subsection{Kaon decay time method}
The second method relies on the measurement of the kaon decay time using
EMC information only for the signal side. We consider events with a $\pi^0$
in the final state:
\begin{equation}
K^\pm \rightarrow X + \pi^0 \rightarrow  X + \gamma \gamma
\end{equation}
In this case the reconstruction efficiency of the kaon decay vertex is
evaluated using a data control sample given by DC information only.
We obtain the kaon decay time using the photon arrival time to the
EMC. From the measured momenta of the $\phi$-meson and of the tagging kaon
we can evaluate the momentum of the signal kaon at the IP and build the
expected path of the signal kaon: it is obtained by geometrical step
by step extrapolation, corrected for local magnetic field and energy loss
effects.  Then we look for clusters not associated to tracks in the DC,
``neutral'' clusters. Among these we select as photon candidates the two
(or three, if present) most energetic clusters with:
\begin{eqnarray}
\nonumber
   && 24^{\circ} < \theta_{\rm{cl}}  < 156^{\circ}\\
   && T_{0} < t_{\rm{cl}} < T_{0} + 70~\rm{ns}
\end{eqnarray}
\noindent
where $\theta_{cl}$ is the polar angle, the angle between the position of
the cluster with respect the IP and the z axis, $t_{cl}$ is the time of the
cluster, and $T_{0}$ is the $\phi$ decay time evaluated considering the
$K_{\mu 2}$ decay chain on the tagging side. These cuts are used to reject
the machine background. The request of three neutral clusters allows us to
select also the events with the track of the charged decay particle not
associated to its calorimeter cluster.\\Then we move along the expected
path of the signal kaon in 5 mm steps and at each step we look for
the $\pi^{0}\to \gamma \gamma$ decay vertex by minimizing a $\chi^2$
function:
\begin{equation}
    \chi^{2} = \frac{(M_{\gamma \gamma} -
      M_{\pi^{0}})^{2}}{\sigma_{M_{\pi^{0}}}^{2}} + \frac{\Delta t_{12}^{2}}
    {\sigma_{t_{12}}^2} + \frac{(t_{K} -
    t_{cl})^{2}}{\sigma_{t}^{2}}
    \label{chi2_nv}
  \end{equation}

In Eq. (\ref{chi2_nv}) $M_{\gamma \gamma}$ is the invariant mass of the two
photon candidates, $M_{\pi^{0}}$ is the $\pi^0$ mass and
$\sigma_{M_{\pi^{0}}}\simeq 17$ MeV is the resolution on the
$\pi^{0}$ mass; $\Delta t_{12}$ is defined as:
  \begin{equation}
    \Delta t_{12} = (t_{\gamma_{1}} - r_{1}/c) - (t_{\gamma_{2}} - r_{2}/c)
    \label{eq:deltat}
  \end{equation}
where $t_{\rm{\gamma_{i}}}$ is the time of the i-th neutral cluster,
$r_{i}$ is the distance of the neutral cluster from the candidate kaon
decay vertex; $\sigma_{t_{12}}$ is obtained by propagating the errors on
the equation \ref{eq:deltat}. $t_{K}$ is the time of flight of the charged
kaon measured along the kaon path, $t_{cl}$ is the time of flight of the
charged kaon given by the weighted mean of the arrival time of the two
photons; $\sigma_{t}$ is the uncertainty on the kaon decay time obtained from
the selected neutral clusters, of the order of few hundreds ps. In case
three clusters have been selected we choose the pair with best
$\chi^2$. The position along the signal kaon path that gives the minimum
value of the $\chi^{2}$ defines the $ K^\pm \rightarrow X + \pi^0
\rightarrow  X + \gamma \gamma $ decay point and is accepted if lying
inside the FV. Moreover we require:
\begin{eqnarray}
\nonumber
  && \chi^{2} < 30,\\
  && 80~\rm{MeV} < M_{\gamma\gamma} < 200~\rm{MeV},\\
  \nonumber
  && \left|\frac{\Delta t_{12}}{\sigma_{t_{12}}}\right| < 5.
\end{eqnarray}
The charged kaon proper time is obtained from:
\begin{equation}
  t^{*} = \frac{t_{cl}}{\langle \gamma \rangle}
  \label{eq:beta_time}
\end{equation}
Where $\langle \gamma\rangle$ is the average between the charged
kaon Lorentz factor $\gamma$ at the IP and at the decay vertex. This
approximation is appropriate because the variation of $\gamma$ along the
decay path is of the order of 1$\%$. The average $t^*$ resolution of the
events selected is better than 1 ns, see fig. \ref{fig:reso_paper}. The
secondary peaks at $\pm$n$\times$2.7 ns are due to events with an incorrect
bunch crossing determination. In the range between 10 and 50 ns we
collected $8\times10^{5}$ events.
%%%%%%%%%%%%%%%%%%%%%%%%%%%%%%%%%%%%%%%%%%%%%%%%%%%%%%%%%%%%
\FIGURE{\figb reso_tau_papero;6.5;\kern1.cm\figb reso_time_papero;6.5;
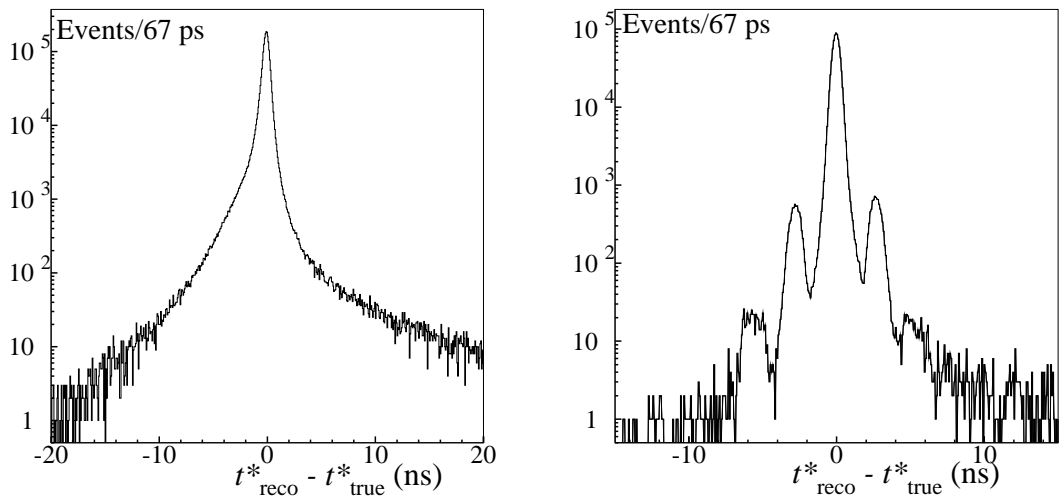
\caption{\footnotesize{$t^*$ resolution from kaon path length
    measurement (left) and from kaon decay time measurement (right). The
    side lobes at $\pm$n$\times$2.7 ns in the right plot are events with
    incorrect bunch crossing time determination. This effect is absent for
    the length measurement.}
\label{fig:reso_paper}}}
%%%%%%%%%%%%%%%%%%%%%%%%%%%%%%%%%%%%%%%%%%%%%%%%%%%%%%%%%%%%%%%
\section{Efficiency evaluation}
\label{sec:eff} In order to be as little dependent as possible
from the MonteCarlo simulation and from possible differences
between data and MonteCarlo, we measured directly on data the
reconstruction efficiency of the kaon decay vertex, as a function
of the charged kaon proper time, for both methods. A data control
sample for the efficiency of the first method, using DC
information for signal selection, is obtained from the signal
selection of the second method, based on EMC information only, and
viceversa. What is relevant for the fit is the efficiency behavior
as a function of $t^*$ rather than its absolute value. In both
cases the efficiency has been obtained in bins of $t^*$ using the
definitions (\ref{eq:beta_time}) and (\ref{eq:beta_ch})
respectively. We applied the same method on MonteCarlo events and
we compared the efficiencies obtained with this method (MC
datalike efficiencies) with the MC true efficiencies, determined
as functions of the true proper time including all the decay modes
used for each method ($K^{\pm} \to all$ for the first method and
$K^{\pm} \to X \pi^0$ for the second method). Fig.
\ref{fig:erika_eff_mc} shows a comparison between the MC true and
the MC datalike efficiencies. For the first method it is important
to stress that even if the control sample is obtained
reconstructing kaon decays with a $\pi^0$ in the final state,
datalike efficiency reproduces the true one obtained
reconstructing all the kaon decay modes. Fig.
\ref{fig:erika_eff_data} shows the ratio of the efficiency
measured on data over the MC datalike efficiency, for both
methods. For the kaon decay length method, the efficiency measured
on data in the range [5, 20] ns is different in shape from the MC
efficiency. The highly ionizing charged kaon fires multiple hits
in the small cells of the DC, i.e. in the inner 12 layers of
detector. This effect has been introduced in the MC simulation
afterwards. For this reason we used the efficiency evaluated
directly data with the small, $\mathcal{O}$(10$^{-4}$), correction
given by the ratio of the MC true efficiencies over MC datalike
efficiencies.
%%%%%%%%%%%%%%%%%%%%%%%%%%%%%%%%%%%%%%%%%%%%%%%%%%%%%%%%%%%%%
\FIGURE{\figb tau_eff_papero;5.5;\kern1.cm\figb
  time_eff_papero;5.5;
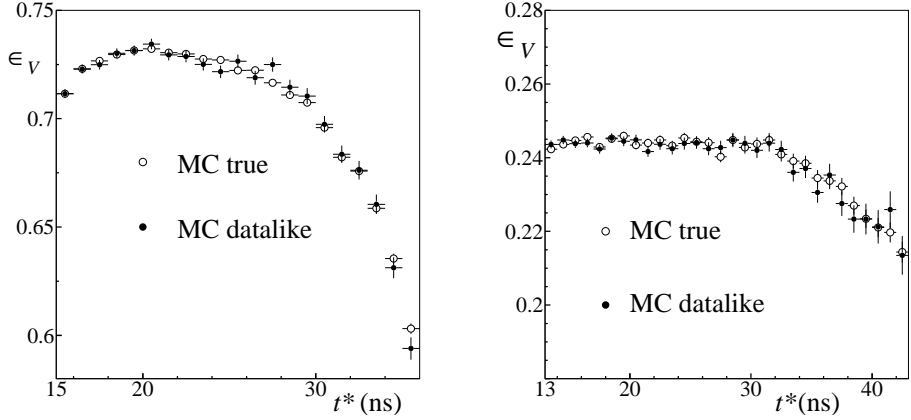
   \caption{Left: charged vertex reconstruction
   efficiency as function of the charged kaon proper time. Right:
   $\pi^0$  decay vertex efficiency as a function of the charged kaon
   proper time. Dots (open circles) represent MC datalike (MC true)
   efficiencies.\label{fig:erika_eff_mc}}}

\FIGURE{\figb ratio_tau_eff_papero;5.5;\kern1.cm\figb
  ratio_time_eff_papero;5.5;
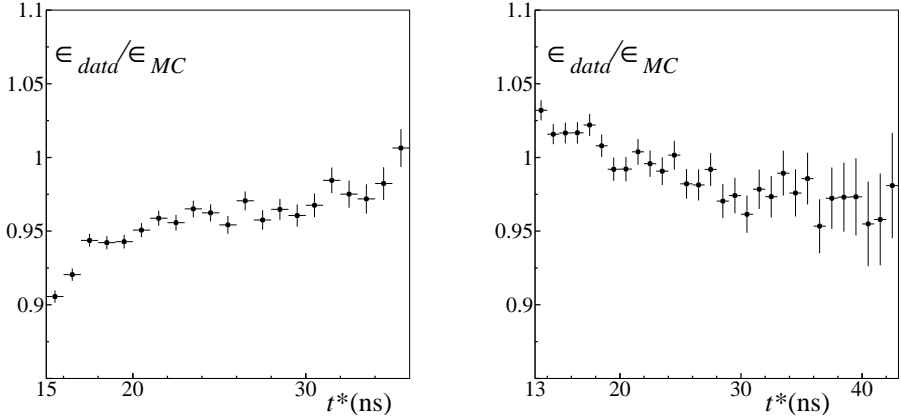
   \caption{Left: ratio between the charged vertex reconstruction
   efficiency as function of the charged kaon proper time evaluated on data
   and the MC datalike one. Right: ratio between the $\pi^0$ decay vertex
   efficiency as a function of the charged kaon proper time evaluated on
   data and the MC datalike one.\label{fig:erika_eff_data}}}
%%%%%%%%%%%%%%%%%%%%%%%%%%%%%%%%%%%%%%%%%%%%%%%%%%%%%%%%%%%%%%%
\section{Fit to the proper time distribution}
\label{sec:fit}
The fit to the proper time distribution is done comparing the observed
with the expected distribution and minimizing a $\chi^{2}$ function. The
entries in each bin of the expected histogram are given by the integral of
the exponential decay function (which depends only on the kaon lifetime)
corrected for the efficiency curve; a smearing matrix accounts for the
effect of the resolution. We fit the proper time distribution in the time
interval where we have good agreement between MC true and MC datalike
efficiencies. The efficiency profiles obtained on data as in
section \ref{sec:eff} are corrected for the tiny residual difference
between the MC true and MC datalike shapes, and for the small
effect of the tag-signal hemispheres correlation; the overall corrections is
of $\mathcal{O}$(10$^{-3}$) for both methods. The resolution function is
evaluated in slices of the order of few nanoseconds on a MC data sample of
about 175 pb$^{-1}$ of $\phi \rightarrow K^+ K^-$ events.
We then define the $N \times N$ smearing matrix,
whose element $s_{ji}$ gives the fraction of events generated in the i-th
bin but reconstructed in the $j-th$ bin. We chose a bin size of 1 ns in order
to reduce the statistical fluctuations and the relative importance of the
smearing corrections. We then minimize
\begin{equation}
\chi^{2} = \sum_{j=1}^{N_{\rm{fit}}}\frac{(N^{\rm{obs}}_{j} -
N^{\rm{ex}}_{j})^{2}}{(\sigma^{\rm{fit}}_{j})^{2}}
\label{chi2_bin}
\end{equation}
where $N^{\rm{obs}}_{j}$ is the number of entries in the $j-th$ bin of the
proper decay time distribution. The content of the $j-th$ bin of
the expected histogram, ($N^{\rm{ex}}_{j}$), is given by:
\begin{equation}
  N^{\rm{ex}}_{j} = \sum_{i = 1}^{N}s_{ji}\times
  \epsilon_{i}\times\epsilon_{i}^{\rm{corr}}\times I_i(\tau),
\label{N_expected}
\end{equation}
where $I_i(\tau)$ is the integral of the exponential over bin $i$,
$\epsilon_{i}$ is the measured reconstruction efficiency and
$\epsilon_{i}^{\rm{corr}}$ is the efficiency correction described above. The
statistical fluctuation of the expected histogram in the bin $j-th$,
$\sigma^{\rm{fit}}_{j}$, is given by the sum of the statistical fluctuation of
the efficiency, the statistical fluctuation of its correction and the
poissonian fluctuation of the expected histogram.
$N_{\rm{fit}}$ is the number of bins used in the fit. The number of bins $N$ of
the expected histogram is larger than $N_{\rm{fit}}$ since we have
allowed the migration from/to the bins used in the fit for a slightly
larger range.\\

\subsection{Kaon decay length results}
We build the expected histogram in the region between 12 and 40 ns
and we fit it in the region between 15 and 35 ns. For $K^+$  we obtained:
\begin{equation}
\tau^+ = (12.338\pm 0.042)~\rm{ns} \hspace{1, cm} \chi^2/\rm{ndf} =
18.2/18 \hspace{1. cm} P(\chi^2) =  44.7\%~.
\end{equation}
While for the $K^{-}$  we obtained:
\begin{equation}
\tau^- =  (12.395\pm 0.045)~\rm{ns} \hspace{1, cm} \chi^2/\rm{ndf} =  26.7/18
\hspace{1, cm} P(\chi^2) =  10.8\%~.
\end{equation}
The left panel of fig. \ref{fig:Data_fit_plus_length} and
\ref{fig:Data_fit_minus_length} shows the data distribution of $t^*$
together with the fit results. The right panel shows the distribution of
the residuals defined as the difference between the data distribution and
the fit. For both charges the distribution of residuals is satisfactory
fitted by a constant compatible with zero within errors.
\FIGURE{\figb tau_plus_papero;6.5;\kern.5cm\figb
  res_tau_plus_papero;6.5;
  \caption{Decay length method. Left: fit to the $K^{+}$
      proper time distribution: the black dots are data
      distribution, in grey the expected histogram
      fitting function. Right: Residuals in the region between 15
      and 35 ns.\label{fig:Data_fit_plus_length}}}
\FIGURE{\figb tau_minus_papero;6.5;\kern.5cm\figb
  res_tau_minus_papero;6.5;
  \caption{Decay length method. Left: fit to the $K^{-}$
      proper time distribution: the black dots are data
      distribution, in grey the expected histogram
      fitting function. Right: Residuals in the region between 15
      and 35 ns.\label{fig:Data_fit_minus_length}}}
The measurements obtained for the two charges are in agreement with each
other. Their weighted mean is:
\begin{equation}
\tau = (12.364\pm 0.031)~\rm{ns}.
\end{equation}
\subsection{Kaon decay time results}
For the second method we build the expected histogram in the region
between 10 and 50 ns. We made the fit in the region between 13 and 42
ns. The result from the fit to the $K^{+}$ proper decay time distribution,
see fig. \ref{fig:Data_fit_plus_time}, is:
\begin{equation}
\tau^+ = (12.315\pm 0.042)~\rm{ns} \hspace{1, cm} \chi^2/\rm{ndf} = 25.5/27
\hspace{1. cm} P(\chi^2) =  49\%.
\end{equation}
For the $K^{-}$ lifetime, see fig. \ref{fig:Data_fit_minus_time},
we obtained:
\begin{equation}
\tau^- =  (12.360\pm 0.043)~\rm{ns} \hspace{1, cm} \chi^2/\rm{ndf} =  22.9/27  \hspace{1, cm}
P(\chi^2) =  69\%.
\end{equation}
\FIGURE{\figb time_plus_papero;6.5;\kern.5cm\figb
  res_time_plus_papero;6.5;
\caption{Decay time method. Left: fit to the $K^{+}$
    proper time distribution: the black dots are data
    distribution, in grey the expected histogram
    fitting function. Right: Residuals in the region between
    13 and 42 ns.\label{fig:Data_fit_plus_time}}}
\FIGURE{\figb time_minus_papero;6.5;\kern.5cm\figb
  res_time_minus_papero;6.5;
\caption{Decay time method. Left: fit to the $K^{-}$
    proper time distribution: the black dots are data
    distribution, in grey the expected histogram
    fitting function. Right: Residuals in the region between
    13 and 42 ns. \label{fig:Data_fit_minus_time}}}
The measurements obtained for the two charges are in agreement with each
other. Their weighted mean is:
\begin{equation}
\tau = (12.337\pm 0.030)~\rm{ns}.
\end{equation}
As shown in the fig. \ref{fig:Data_fit_plus_length},
\ref{fig:Data_fit_minus_length}, \ref{fig:Data_fit_plus_time} and
\ref{fig:Data_fit_minus_time} the agreement between data distribution and
$\it{expected}$ histo distribution is very good also outside the fit range
for both the charges and both the methods.
%%%%%%%%%%%%%%%%%%%%%%%%%%%%%%%%%%%%%%%%%%%%%%%%%%%%%%%%%%%%%%%
\section{Systematic uncertainties}
\label{Systematics}
All the measurement strategy has been designed in order to have redundancy
to keep systematic effects under control, and also to use as much as
possible control samples selected on data, rather than relying on
MonteCarlo simulations, for all the inputs to the fit. The determination of
efficiencies has been already discussed; the efficiency correction effect
has been checked by performing the fit with and without it.
 For the resolution functions,
which at first order we obtain from MonteCarlo, we have checked
the data/MC agreement on the distribution of the difference
between the $t^*$ from the decay length and the $t^*$ from the
decay time, on doubly reconstructed events. This allows us to
check also the amount of incorrect bunch crossing determinations
since the length is clearly not affected by this feature. The core
of the time resolution is also kept under control by comparing the
time of the two neutral clusters. All checks have shown very small
data/MC corrections and negligible variations to the fit results
and quality. The stability of the fit procedure with respect to
the fit range has been checked by varying significantly the fit
window and by changing the bin size from 1 ns to 0.5 or 2 ns. The
systematics due to the selection have been obtained by comparing
the fit results with different cuts for both the signal and the
control samples. The beam pipe and DC wall thickness are known to
about 10$\%$; this in turn affects the evaluation of the kaon
energy loss in the materials and consequently the kaon $\beta$. We
have repeated both measurements considering different thicknesses
(by varying them within their errors) in order to assess the
systematic effect on the lifetime. The energy loss inside the DC
is more than one order of magnitude smaller than the one in the
walls and the uncertainty on the material crossed inside the DC
gives a negligible contribution to the systematic error on the
lifetime measurements. The first method is very sensitive to
systematic shift on the $\beta$ of the kaon (see eq.
\ref{eq:beta_ch}) and we observe a sizeable effect on the
lifetime. This is not true for the second method (see eq.
\ref{eq:beta_time}): for a typical charged kaon in our detector
$\delta\gamma/\gamma \simeq 0.06 \times \delta \beta/\beta$ and
$t_{cl}$ depends on the kaon energy loss only at second order, via
the third term of the $\chi^2$ function in eq. \ref{chi2_nv}. For
the time method the overall systematic effect due to the
uncertainty on the kaon energy loss has been actually found
negligible. The estimates of the systematic effects are summarized
in table \ref{tab:tot_sys}. The systematics of the two
measurements are almost completely uncorrelated. \noindent \TABLE{
\begin{tabular}[c]{|c|c|c|}
\hline
 \bf{systematic uncertainty} & \bf{Length (ps)} & \bf{Time (ps)} \\
\hline
Fit range & $\pm$ 10 & $\pm$ 10\\
\hline
binning & $\pm$ 10 & $\pm$ 10\\
\hline
efficiency correction & $\pm$ 15 & $\pm$ 10\\
\hline
beam pipe thickness & $\pm$ 10 & negligible\\
\hline
DC wall thickness & $\pm$ 15 & negligible\\
\hline
Signal selection cuts & $\pm$ 15 & negligible\\
\hline
Control sample cuts & negligible & $\pm$10\\
\hline
Resolution effects & negligible & negligible\\
\hline
total systematic uncertainty & $\pm$ 31 & $\pm$ 20\\
\hline
\end{tabular}
  \caption{\footnotesize{Sources of systematic uncertainties.}}
\label{tab:tot_sys}}
The result from the length measurement is:
\begin{equation}
\tau = (12.364\pm 0.031_{\rm{stat}}\pm 0.031_{\rm{syst}})~\rm{ns}
\end{equation}
and from the time measurement is:
\begin{equation}
\tau = (12.337\pm 0.030_{\rm{stat}}\pm 0.020_{\rm{syst}})~\rm{ns}
\end{equation}

%%%%%%%%%%%%%%%%%%%%%%%%%%%%%%%%%%%%%%%%%%%%%%%%%%%%%%%%%%%%%%%
\section{Correlation and average}
\label{sec:crl}
In order to average the two methods we calculate the statistical
correlation between the two results. The correlation arises because the
data samples used for the two methods have $\approx 30 \%$ of events in
common. The normalized correlation is 30.7$\%$ in agreement with a
direct estimate obtained dividing the data in subsamples. Assuming that the
systematic uncertainties are uncorrelated we then obtain the average:
\begin{equation}
\tau = (12.347\pm 0.030)~\rm{ns}
\end{equation}
The measurement obtained agrees, within the errors, with the result given by
Ott and Pritchard, \cite{ott}:
\begin{equation}
\tau = (12.380\pm 0.016)~\rm{ns}
\end{equation}
 and with the PDG fit, \cite{PDG}:
\begin{equation}
\tau = (12.384\pm 0.024)~\rm{ns}
\end{equation}
%%%%%%%%%%%%%%%%%%%%%%%%%%%%%%%%%%%%%%%%%%%%%%%%%%%%%%%%%%%%%%%
\section{CPT test}
The comparison of $K^+$ and $K^-$ lifetimes is a test of $CPT$ invariance
which requires the equality of the decay lifetimes for
particle and antiparticle. The average of the two methods is:
\begin{equation}
\begin{split}
%\begin{eqnarray}
 \tau^+ = (12.325 \pm 0.038 )~\rm{ns}\\
 \tau^- = (12.374 \pm 0.040 )~\rm{ns}
%\end{eqnarray}
\end{split}
\end{equation}
From these measurements, and taking into account that most of the
systematic effects cancel out in the ratio, we obtain:
\begin{equation}
\frac{\tau^-}{\tau^+} = 1.004 \pm 0.004
\end{equation}
Our result agrees with $CPT$ invariance at the 0.4$\%$ level. Ref
\cite{lobkowicz} had already verified agreement at the 0.08$\%$ level.
%\newpage
\section*{Acknowledgments}
We thank the DAFNE team for their efforts in maintaining low background
running conditions and their collaboration during all data-taking.
We want to thank our technical staff: G.F.Fortugno and F.Sborzacchi for
their dedicated work to ensure an efficient operation of the KLOE Computing
Center; M.Anelli for his continuous support to the gas system and the safety of
the detector; A.Balla, M.Gatta, G.Corradi and G.Papalino for the
maintenance of the electronics; M.Santoni, G.Paoluzzi and R.Rosellini for
the general support to the detector; C.Piscitelli for his help during major
maintenance periods. This work was supported in part by EURODAPHNE,
contract FMRX-CT98-0169; by the German Federal Ministry of Education and
Research (BMBF) contract 06-KA-957; by the German Research Foundation
(DFG), 'Emmy Noether Programme', contracts DE839/1-4; by INTAS, contracts
96-624, 99-37; and by the EU Integrated Infrastructure Initiative
HadronPhysics Project under contract number RII3-CT-2004-506078.
\bibliographystyle{elsart-num}
\bibliography{biblio}
\end{document}